%% file: ms.tex
\documentclass[letterpaper, 10 pt, conference]{ieeeconf}  
\IEEEoverridecommandlockouts        % This command is only needed if you want to use the \thanks command
\usepackage{microtype}
\usepackage{times} % assumes new font selection scheme installed
\usepackage{amsmath} % assumes amsmath package installed
\usepackage{amssymb}  % assumes amsmath package installed
\usepackage{multirow}
\usepackage[compress,sort]{cite}

\usepackage[hidelinks,pdftex,pdfauthor={Justin Miller, Jonathan P.\ How},pdftitle={Predictive Positioning and Quality Of Service Ridesharing for Campus Mobility On Demand Systems}]{hyperref}

\title{\LARGE \bf Predictive Positioning and Quality Of Service Ridesharing for Campus Mobility On Demand Systems}

\author{Justin Miller and Jonathan P.\ How% <-this % stops a space
  \thanks{Laboratory of Information and Decision Systems,
    Massachusetts Institute of Technology, 77 Massachusetts Ave.,
    Cambridge, MA, USA {\tt\small \{justinm, jhow\}@mit.edu}}%
}

\usepackage{mathtools}

\usepackage[labelfont=bf,font=small]{caption}
\usepackage{float}
\usepackage[labelfont=bf,font=small]{subcaption}

\usepackage{amssymb}

\usepackage[linesnumbered,ruled,vlined]{algorithm2e}

\DeclareMathOperator*{\argmin}{arg\,\!min}

\usepackage{balance}

\usepackage[capitalize]{cleveref}
\crefformat{equation}{(#1)}
\Crefformat{equation}{(#1)}
\crefname{equation}{}{}
\Crefname{equation}{}{}
\Crefname{table}{Table}{Tables}
\Crefname{figure}{Figure}{Figures}

\allowdisplaybreaks
\makeatletter
\renewcommand\paragraph{\@startsection{subsubsection}{4}{\z@}%
	{0.5ex \@plus1ex \@minus.2ex}%
	{-.15em}%
	{\normalfont\normalsize\itshape}}
\makeatother

\begin{document}

\maketitle
\thispagestyle{empty}
\pagestyle{empty}

%%%%%%%%%%%%%%%%%%%%%%%%%%%%%%%%%%%%%%%%%%%%%%%%%%%%%%%%%%%%%%%%%%%%%%%%%%%%%%%%
\input{abstract}

%%%%%%%%%%%%%%%%%%%%%%%%%%%%%%%%%%%%%%%%%%%%%%%%%%%%%%%%%%%%%%%%%%%%%%%%%%%%%%%%
\section{Introduction}\label{sec:introduction}
\input{introduction}

%\section{Related Work}\label{sec:related_work}
%\input{related_work}

\section{Problem Formulation}\label{sec:problem_formulation}
\input{problem_formulation_arxiv}

\section{Approach}\label{sec:approach}
\input{approach}

\section{Experiments}\label{sec:experiments}
\input{experiments_arxiv}

\section{Conclusion}\label{sec:conclusion}
 \input{conclusion}

\section*{Acknowledgment}
Research supported by the Ford Motor Company through the Ford-MIT Alliance.

\balance

\bibliographystyle{IEEEtran}
\bibliography{ICRA2017}
%%%%%%%%%%%%%%%%%%%%%%%%%%%%%%%%%%%%%%%%%%%%%%%%%%%%%%%%%%%%%%%%%%%%%%%%%%%%%%%%

\pagebreak
\appendix
\input{appendix}

\addtolength{\textheight}{-0cm}
\end{document}

%% file: abstract.tex
\begin{abstract}
Autonomous Mobility On Demand (MOD) systems can utilize fleet management strategies in order to provide a high customer quality of service (QoS).
Previous works on autonomous MOD systems have developed methods for rebalancing single capacity vehicles, where QoS is maintained through large fleet sizing.
This work focuses on MOD systems utilizing a small number of vehicles, such as those found on a campus, where additional vehicles cannot be introduced as demand for rides increases.
A predictive positioning method is presented for improving customer QoS by identifying key locations to position the fleet in order to minimize expected customer wait time.
Ridesharing is introduced as a means for improving customer QoS as arrival rates increase.
However, with ridesharing perceived QoS is dependent on an often unknown customer preference.
To address this challenge, a customer ratings model, which learns customer preference from a 5-star rating, is developed and incorporated directly into a ridesharing algorithm.
The predictive positioning and ridesharing methods are applied to simulation of a real-world campus MOD system.
A combined predictive positioning and ridesharing approach is  shown to reduce customer service times by up to 29\%. and the customer ratings model is shown to provide the best overall MOD fleet management performance over a range of customer preferences. 
\end{abstract}

%% file: introduction.tex
Mobility On Demand (MOD) systems have the potential to revolutionize transportation systems in urban settings by providing commuters access to vehicles without requiring private ownership.
In such systems, a fleet of shared vehicles continually services multiple customers by transporting them from their requested on demand pickup location to their desired destination.
It is estimated that by 2030, as much as 26\% of all global miles traveled will be from customers using shared vehicles \cite{_morgan_2016}.
A fundamental challenge for MOD systems is providing a high customer \textit{quality of service} (QoS) in order to minimize any drawbacks that customers may experience by relying on the shared resources. 

\begin{figure}[t]
	\centering
	\begin{subfigure}[t]{0.335\textwidth}
	\includegraphics[width=1\columnwidth]{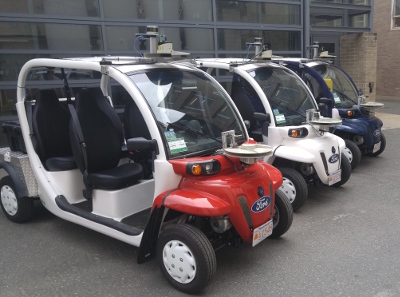}
	\caption{MIT MOD fleet}
	\label{fig::fleet}
	\end{subfigure}
	\begin{subfigure}[t]{0.14\textwidth}
	\includegraphics[width=1\columnwidth]{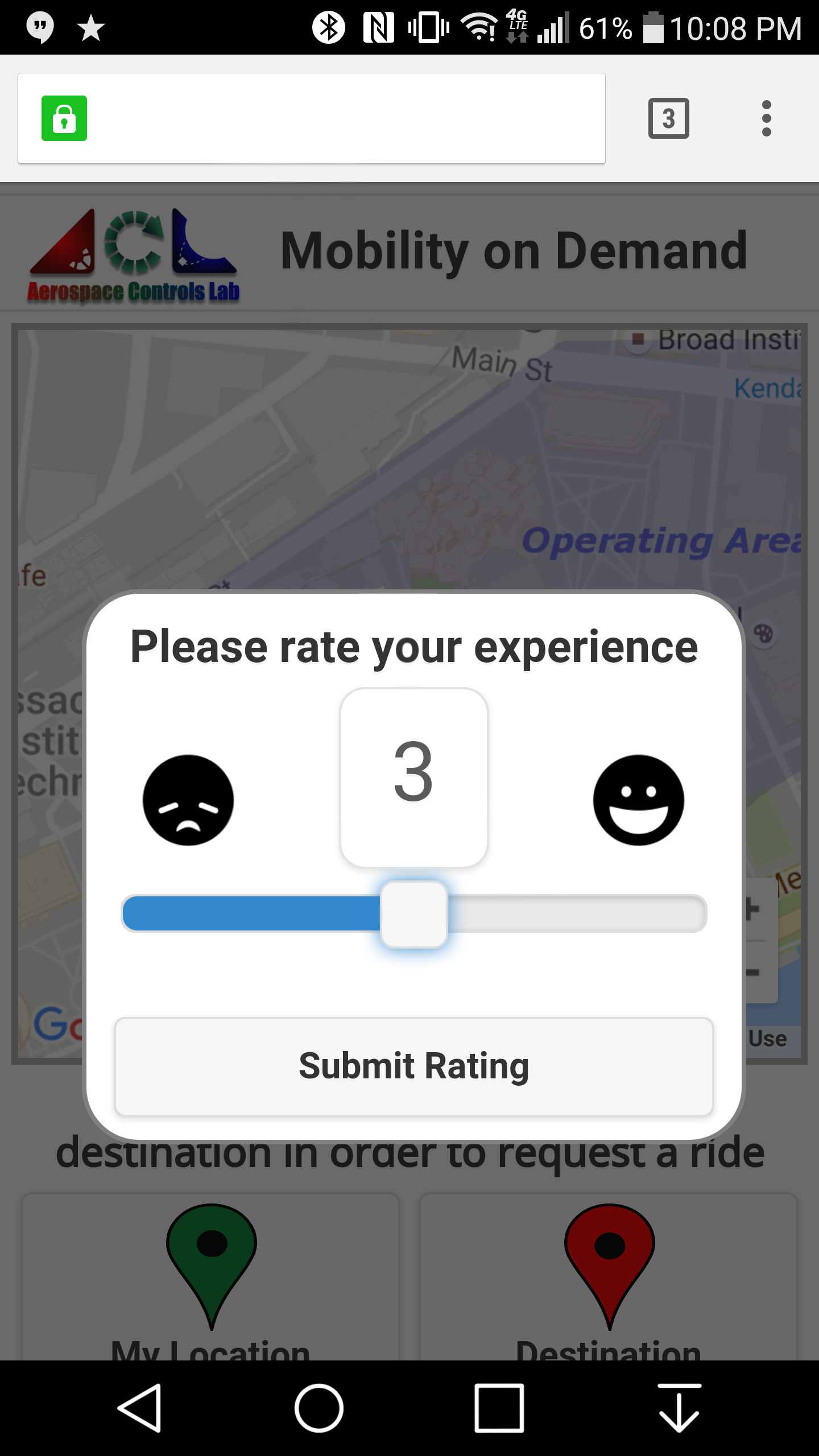}
	\caption{Rating app}
	\label{fig::app}
	\end{subfigure}
	\caption{(a) A fleet of three electric shuttles operate as MOD vehicles on MIT campus. (b) MIT MOD customers have the ability to provide app-based feedback in the form of a 5-star rating.}
	\label{fig::fleet_and_app}
\end{figure}

%WHAT IS OUR APP
%VISION FOR APP
There are many factors that can affect a customer's QoS such as cost, comfort, safety and convenience \cite{paquette_quality_2009}, many of which can be improved through the use of autonomous MOD systems composed of self-driving vehicles \cite{pavone_autonomous_2015}.
Several autonomous MOD fleet management strategies have been introduced as a means of improving QoS.
These approaches attempt to find a ``rebalancing" policy that redistributes vehicles within the system based on customer demand using either a fluid model approach \cite{pavone_robotic_2012}, a Markov transition model approach \cite{volkov_markov-based_2012}, a queueing-theoretical model approach \cite{zhang_control_2016-1}, or a model predictive control approach \cite{zhang_model_2016}.
The approaches were developed to operate on city-wide scales and assume that an appropriately large fleet of autonomous vehicles is available, an assumption that does not yet reflect the current state of real-world autonomous MOD systems.
In practice, autonomous MOD systems are being deployed with limited sized fleets \cite{abuelsamid_singapore_????,ross_helsinki_2016,ross_uber_2016}.
This work is further motivated by our own MOD system operating on MIT campus, where autonomous fleet management strategies are applied to a fleet of only three human-driven vehicles, shown in \cref{fig::fleet}.
Previous rebalancing methods do not scale well when applied to such small fleets.
For example, the policies are designed to redistribute multiple excess vehicles to relatively few network arrival locations. 
The methods breakdown when the reverse is true and fleet sizes are relatively small compared to the number of network locations such that there generally aren't enough vehicles to cover all arrival locations.
This work seeks to address this challenge by instead identifying key locations in the MOD system which minimize expected customer wait times regardless of fleet size.

Additionally, those previous works only consider the use of single capacity vehicles and do not address the benefits and challenges of utilizing ridesharing.
In ridesharing, multiple customers may share a ride in an MOD vehicle at the same time.
Newly arrived passengers can be picked up before onboard passengers have been dropped off, allowing for more customers to be serviced with fewer vehicles.
However, the reduced wait time for requested passengers comes at the expense of increased ride time of onboard passengers. 
With ridesharing, perceived QoS becomes dependent on customer preference (i.e. how much customers prefer one service metric over another).
A large body of work has studied a form of ridesharing known as the Dial-A-Ride Problem (DARP), which is a specialization of the Vehicle Routing Problem formulated specifically for the transportation of customers.
DARP problem formulations typically take either the form of an integer program or a scheduling problem.
Integer program formulations encode each customer QoS metric as a decision variable and use heuristic methods such as genetic algorithms \cite{jorgensen_solving_2007}, simulated annealing \cite{mauri_customers_2009}, and tabu search \cite{paquette_combining_2013} to minimize an objective, such as a cost function composed of a weighted sum of the metrics.
Scheduling problem formulations enumerate the possible ways of inserting new passengers into vehicle schedules, and encode customer QoS metrics as feasibility constraints \cite{jaw_heuristic_1986,coslovich_two-phase_2006,fagnant_dynamic_2015-1}.
The main challenge with all of these approaches is that the weights or constraint thresholds that define customer preference may be chosen incorrectly and even the structure that encodes the QoS metrics could be wrong.
In general determining customer preference can be difficult.
However, many MOD systems such as Uber, Lyft, and the MIT MOD system are able to query passengers for feedback in the form of a 5-star rating using a ride request app, as shown in \cref{fig::app}.
This work takes advantage of this available information through a ridesharing algorithm which does not encode customer preference directly, but rather utilizes a learned customer ratings model when solving the DARP scheduling problem.

%WHAT ARE THE CONTRIBUTIONS
The contributions of this work are: 1) a predictive positioning approach that minimizes expected customer wait time in MOD networks with fewer vehicles than customer arrival locations; 2) a customer rating model which learns from 5-star rating feedback and serves as a customer QoS-focused cost function; and 3) a schedule-based ridesharing framework that accommodates customer QoS metrics without the need for encoding customer preference constraints.

%% file: problem_formulation_arxiv.tex
\paragraph{Customer Arrival Model}
An MOD network is modeled as a directed \textit{network} graph denoted by $\mathcal{G} = (\mathcal{N},\mathcal{L})$, where $\mathcal{N} = \{ n_1, \dots, n_{N_n} \}$ is a set of $N_n$ nodes, and $\mathcal{L} = \{l_1, \dots, l_{N_l}\}$ is a set of $N_l$ directed link edges each taking the form of an ordered pair of neighbor nodes $l = (n_i, n_j) \in \mathcal{N}^2$.
A route $r(n_i,n_j)$ is defined as a sequence of directed links $\mathcal{L}_{r} \subseteq \mathcal{L}$ which corresponds to a unique minimum-travel-time path between a pair of nodes $(n_i, n_j)$.
Customers are assumed to arrive at nodes in the network graph according to a Poisson process with arrival rate parameter $\lambda_n$.
In many MOD systems, customer arrival rates will be time-varying and may feature large fluctuations on short-time scales.
In this work, a discrete-time approximation is used where Poisson arrival rates are static for the operating duration, with short-term fluctuations averaged out across the time period.

\paragraph{Ridesharing}
Upon arrival, customers send ride requests consisting of a pickup node $p \in \mathcal{N}$ and a drop off node $e \in \mathcal{N}$.
Let $\mathcal{C}$ of size $N_c$ be the set of customers who have requested rides, let $\mathcal{O}$ of size $N_o = 2N_c$ be the set of requested customer pickup and drop off nodes, and let $\mathcal{V}$ of size $N_v$ be the set of vehicles in the MOD fleet. 
Vehicle $v$ will service a subset of the customers $\mathcal{C}_v \subseteq \mathcal{C}$ by visiting their pickup and drop off nodes.
All customer nodes are inserted into a schedule $\mathbf{s_v} = \{s_1, \dots s_{N_o}\}$, $s_i~\in~\{\emptyset,\mathcal{N}\}$, which the vehicle traverses in order using a sequence of routes. 
Let $Q$ be the maximum capacity of each vehicle.
The ridesharing problem is that of finding both the customer to vehicle assignments as well as the vehicle schedules such that the cost of scheduling all customers is minimized.
A four-index ILP formulation for the DARP is proposed as an extension of the models presented in \cite{cordeau_dial--ride_2007}, with emphasis placed on the ordering of customers within schedules.
The decision variables $x_{cij}^v \in \{0,1\}$ are equal to 1 if vehicle $v$ is assigned customer $c$, with $p_c$ and $e_c$ respectively sequenced at $s_i$ and $s_j$; and zero otherwise.
The objective of the ridesharing problem formulation is to minimize the total customer QoS cost, that is,
\begin{flalign}
\argmin_{x_{cij}^v}  &\quad\sum_{v \in V} \sum_{c \in C} \sum_{i=1}^{N_o} \sum_{j=1}^{N_o} g_{cij}^v x_{cij}^v \label{eqn::objective}\\
s.t.  &\sum_{v \in V}\sum_{i=1}^{N_o} \sum_{j=1}^{N_o} x_{cij}^v = 1 
\quad\quad\quad\quad\quad\forall c \in C  
\label{eqn::customers_assigned}\\
&\sum_{i=1}^{N_o}\sum_{j=1}^{i} x_{cij}^v = 0 
\quad\quad\quad\quad\quad\forall v \in V, c \in C
\label{eqn::pickup_first}\\
&\sum_{c \in C} x_{cij}^v = 1 
\quad\quad\forall \, v \in V, \; i,j \in \{1,\dots,N_o\} 
\label{eqn::schedule_location}\\
&\sum_{c \in C}\sum_{i=1}^{k}\sum_{j=k+1}^{N_o} x_{cij}^v \le Q 
\quad\forall v \in V, \; k \in \{1,\dots,N_o\}
\label{eqn::capacity}\\
&x_{cij}^v \in \{0,1\} 
\quad\quad\forall v \in V, c \in C, i, j \in \{1,\dots,N_o\} \nonumber,
\end{flalign}
where \cref{eqn::customers_assigned} enforces that all customers are assigned, \cref{eqn::pickup_first} enforces that a customer is picked up before being dropped off, \cref{eqn::schedule_location} enforces that only one customer can occupy a given schedule position, and \cref{eqn::capacity} enforces that the capacity of the vehicle is not exceeded.
The cost, further described in \Cref{sec::cost_function}, is chosen to be a function of the QoS metrics $\mathbf{m}_c$ that a customer experiences, that is $g_{cij}^v = g(\mathbf{m}_c^{vij})$ where the metrics themselves are a function of the customer's position in a vehicle's schedule.

\paragraph{QoS Metrics} \label{sec::qos_metrics}
Customer QoS is quantified using a set of transportation metrics, similar to those in \cite{paquette_quality_2009}.
First, customer requests are either accepted by the MOD system, in which case the customers will be assigned to a vehicle and given a ride, or rejected by the system and the customers will walk.
Let $1^\text{rej}$ be a rejection indicator variable which takes value 1 if the customer is rejected and 0 otherwise.
The remaining customer QoS metrics are: ride time $t^{\text{ride}}$, wait time $t^{\text{wait}}$, service time $t^{\text{service}}$, ratio of ride time to direct time $t^\text{ratio}$, excess ride time $t^\text{excess\_ride}$, maximum number of stops while user is onboard $N^{\text{stops}}$, the notification time $t^{\text{notify}}$, the total traveled distance while onboard $d^{\text{traveled}}$, the time it would have taken the customer to walk, $t^{\text{walk}}$, and service time in excess of walk time $t^{\text{excess\_walk}}$.
Let $\mathbf{m}_c \in \mathbb{R}^{11}$ be the QoS metrics for customer $c$, that is $\mathbf{m}_c = \{1^\text{rej}_c, t^{\text{ride}}_c, t^{\text{wait}}_c, t^{\text{service}}_c, t^{\text{ratio}}_c, t^{\text{excess\_ride}}_c, N^{\text{stops}}_c, \allowbreak t^{\text{notify}}_c, d^{\text{traveled}}_c, t^{\text{walk}}, t^{\text{excess\_walk}}_c \}$.
Evaluation of these metrics is further detailed in \cref{sec::ride_metrics}.

%% file: approach.tex
This section present two new approaches for utilizing and solving the ridesharing problem, \cref{eqn::objective,eqn::customers_assigned,eqn::pickup_first,eqn::schedule_location,eqn::capacity}, in order to improve customer service under different operating regimes.
The first approach manages vehicles in the absence of ride requests by using a \textit{predictive positioning algorithm}, while the second approach manages vehicles to accommodate ride requests using a \textit{ridesharing algorithm}.
Additionally, two QoS focused cost functions are presented, a traditional weighted cost function, and a customer rating model which learns customer QoS preference from customer rating feedback.

\subsection{Predictive Positioning}
The predictive positioning algorithm uses known customer arrival rates $\{\lambda_1, \dots \lambda_{N_n}\}$ to find the key \textit{predictive nodes} within the network graph to place unassigned vehicles such that the expected wait time for arriving customers is minimized.
A special form of the ridesharing problem, \cref{eqn::objective,eqn::customers_assigned,eqn::pickup_first,eqn::schedule_location,eqn::capacity}, is solved for a sequence of $N_a$ predicted customer arrivals, where 1) vehicles are only assigned to at most one predicted customer ($Q=1$);  2) the number of considered customers is set to be the number of vehicles ($N_a = N_v$); and 3) the cost is set to be the customer wait time.
Let the vector $\mathbf{k}~\in~\{0,1,\dots,N_v\}^{N_n}$ denote the number of vehicles located at each node and $\mathbf{a}~\in~\{0,1,\dots,N_a\}^{N_n}$ be the number of customer arrivals on each node.
$\mathcal{K}~{=\{\mathbf{k} \mid \sum_{i=1}^{N_n} k_{i}=N_v\}}$ is the set of all possible vehicle placement options, and $\mathcal{A} ~{= \{\mathbf{a} \mid \sum_{i=1}^{N_n} a_{i}=N_a \}}$ is the set of all combinations of possible arrivals.
The predictive node locations $\mathbf{k}^*$ for which to place vehicles is determined using \cref{alg::predictive_positioning}.

\begin{algorithm}[t]
	\textbf{Input:} customer node arrival rates $\{\lambda_1, \dots \lambda_{N_n}\}$\\
	\textbf{Output:} vehicle locations $\mathbf{k}^*$ \\
	enumerate vehicle placement options $\mathcal{K}$ \\
	enumerate possible arrival locations $\mathcal{A}$ \\
	\For{ $\mathbf{a} \in \mathcal{A}$ }{ 
		\For{ $\mathbf{k} \in \mathcal{K} $ }{ 
			$w_{k,a}$ $\leftarrow$ computeTotalWaitTime($\mathbf{k},\mathbf{a}$) \label{eqn::wait_time_cost}\\
		}
		$p_a$ $\leftarrow$ computeProbability($\mathbf{a}$) \label{eqn::probability}\\
	}
	$\mathbf{k}^* = \argmin\limits_{\mathbf{k} \in \mathcal{K}} \sum\limits_{\mathbf{a}\in\mathcal{A}} w_{k,a} p_a$\\
	\Return $\mathbf{k}^*$
	\caption{Predictive Positioning}
	\label{alg::predictive_positioning}
\end{algorithm}

The wait time cost $w_{k,a}$ in \cref{eqn::wait_time_cost} is determined by using a greedy solution to the ridesharing problem. 
The wait times are assumed to be dependent only on the structure of the network graph, and therefore are computed and stored offline.
To handle the cases where some vehicles are serving customers while others need to be predictively positioned, wait times are computed for all numbers of free vehicles from 1 to $N_v$.

The probability of a set of arrivals $p_a$ in \cref{eqn::probability}  is determined using decomposition of the total network arrival Poisson process.
Given that an arrival occurs, the probability of that arrival occurring at a node $n_i$ is given by $P(a_i=1 \mid N_a=1) = \frac{\lambda_{n_i}}{\Lambda}$.
The probability of a set of arrivals occurring according to $\mathbf{a}$ follows a multinomial distribution,
\begin{equation}
P(\mathbf{a}) =
\frac{{N_a}!}{a_1! \dots a_{N_n}!} \left(\frac{\lambda_{1}}{\Lambda}\right)^{a_1} \cdots \left(\frac{\lambda_{N_n}}{\Lambda}\right)^{a_{N_n}}.
\end{equation}
Arrival probabilities are computed online whenever customer arrival rates change.

\subsection{Ridesharing}
\subsubsection{Ridesharing Algorithm}

\begin{algorithm}[t]
	\textbf{Input:} request $\{p_c,e_c\}$, previous customer allocations $\{\mathcal{C}_1,\dots \mathcal{C}_{N_v}\}$, schedules $\{\mathbf{s}_1, \dots, \mathbf{s}_{N_v}\}$\\
	\textbf{Output:} schedule for assigned vehicle $s_{v^*}$\\
	\For{ $v=1:N_v$ }{ 
		$\hat{\mathcal{C}}_v \leftarrow \{c,\mathcal{C}_v\}$ \label{eqn::assign_customer}\\
		$\hat{\mathcal{S}}_v \leftarrow$ enumerateInsertions($\mathbf{s}_v,p_c,e_c$) \label{(eqn::enumerate_insertions)}\\
		$\mathbf{s}^*_v = \argmin\limits_{\hat{\mathbf{s}}_v \in \hat{\mathcal{S}}_v} \sum\limits_{c \in \hat{\mathcal{C}}_v} g(h(\hat{\mathbf{s}}_v, i_c, j_c )) $ \label{eqn::best_schedule}\\
		$\mathcal{M}^*_v \leftarrow$ computeNewMetrics($\mathbf{s}^*_v $)\label{eqn::best_metrics}\\
		$\mathcal{M}_v \leftarrow$ computeBaselineMetrics($\mathbf{s}_v $)\label{eqn::baseline_metrics}
	}
	$\mathcal{M}^*_{N_v+1} \leftarrow $ computeRejectionMetrics($p_c,e_c$) \label{eqn::rejection_metrics}\\
	$v^* \leftarrow $ assignToLowestBid($\{\mathcal{M},\mathcal{M}^*\}_{1:N_v},\mathcal{M}^*_{N_v+1}$) \label{eqn::assigned_vehicle}\\
	$\mathbf{s}_{v^*} \leftarrow \mathbf{s}^*_{v^*}$ \label{eqn::update_schedule}\\
	\Return $\mathbf{s}_{v^*}$
	\caption{Ridesharing}
	\label{alg::ridesharing}
\end{algorithm} 

The ridesharing algorithm uses an insertion method as a heuristic solution to the ridesharing problem.
\Cref{alg::ridesharing} presents the method for assigning a new customer request to a vehicle such that the total QoS cost to the system is minimized.
The algorithm is executed online whenever a new customer ride request is submitted.
\Cref{eqn::assign_customer} temporarily assigns the new customer to each vehicle generating a temporary customer allocation $\hat{C}_v$.
\Cref{(eqn::enumerate_insertions)} enumerates all feasible ways of inserting the request into the schedule, where $\hat{\mathcal{S}}_v$ is the set of all feasible schedules $\hat{\mathbf{s}}_v$, and where feasibility is met by ensuring that $p_c$ is inserted before $e_c$ and that vehicle capacity is not exceeded.
\Cref{eqn::best_schedule,eqn::best_metrics} find the best feasible schedule $\mathbf{s}^*_v$ and corresponding new customer QoS metrics $\mathcal{M}^*_v$ for each vehicle, where $\mathcal{M}^*_v = \{\mathbf{m}_{\bar{c}} \mid \bar{c} \in \mathcal{\hat{C}}_v \}$.
For comparison, \cref{eqn::baseline_metrics} computes the original customer QoS metrics $\mathcal{M}_v$ for each vehicle, where $\mathcal{M}_v = \{\mathbf{m}_{\bar{c}} \mid \bar{c} \in \mathcal{C}_v \}$.

Rather than imposing constraints on customer QoS metrics, the algorithm uses a virtual ``rejection vehicle", $v = N_v+1$ to make bids that consider the case where the customer is not serviced by the MOD system.
\Cref{eqn::rejection_metrics} computes the rejection vehicle customer metrics $\mathcal{M}^*_{N_v+1} = \mathbf{m}_c^\text{rejected}$.
While seemingly counter-intuitive, rejections are in fact important for improving customer QoS. 
For example, if a customer's wait time is significantly longer than the time it would take for them to walk, then they may prefer to be rejected rather than to wait to use the service.
In this work, rejected customers are prevented from making additional requests, although this could be adapted to allow customers to resubmit with QoS relaxations.

\Cref{eqn::assigned_vehicle} finds the combination of baseline and new metrics that includes all customers and has the lowest overall customer QoS cost, and then returns the vehicle $v^*$ that contains $c$.
If $v^*$ is the rejection vehicle, then the customer is rejected, otherwise the schedule for $v^*$ is updated to accommodate the request.

The primary benefit of the ridesharing algorithm is the ability to evaluate customer QoS without having to encode the customer preference structure into the algorithm.
For example, other methods \cite{jaw_heuristic_1986,coslovich_two-phase_2006,fagnant_dynamic_2015-1} encode feasibility constraints on customer metrics such as wait time or ride time, where customers are rejected if these are not met. 
Instead, a more general approach is taken in \cref{alg::ridesharing} where a competing bid is made to reject a customer, and the rejection is made only when the overall QoS of the system would be improved by doing so.
This approach opens the door for a ratings based cost function where bids are made without constraining the customer metrics directly.

\subsubsection{Ridesharing Cost Functions} \label{sec::cost_function}
Two ridesharing cost functions are presented which evaluate the customer QoS cost from a set of customer ride metrics.
First, a cost function composed of a linear weighted combination of the customer metrics is proposed as
\begin{equation}
g(\mathbf{m}_c) = \sum_{i=2}^{|m_c|} w^\text{rej}_i  1^\text{rej}_c  m_{c,i} + w^\text{acpt}_i  (1-1^\text{rej}_c)  m_{c,i} ,
\label{eqn::linear_weight_cost_functoin}
\end{equation}
where $w^\text{rej}_i$ and $w^\text{acpt}_i$ are weights for metric $i$ that are used to allow for differentiating between rejected and serviced customers.
For example, a service time focused cost function would be $g(\mathbf{m}_c) = 1^\text{rej}_c t^{\text{walk}}_c  + (1 - 1^\text{rej}_c) t^{\text{service}}_c$, where the cost is the service time if the customer receives a ride and the walk time if they are rejected.
This form of cost function requires that the weights be properly chosen to reflect customer preference, and can result in poor customer QoS if the weights are wrongly chosen.

To overcome the need to choose weights, a second ratings based cost function is presented which learns and uses customer preference through feedback from 5-star ratings. 
The rating model utilizes a random forest of classification decision trees, based on the work of \cite{breiman_random_2001}.
Random forest algorithms tend to prevent overfitting and have been demonstrated to perform well empirically \cite{caruana_empirical_2006}.
To train the random forest model, a dataset $D$ from $N_D$ customers is collected in the form of 5-star ratings $Y_\text{train} = \{y_1, \dots, y_{N_D}\}$ and a ride metrics feature vector $X_\text{train} = \{m_1, \dots, m_{N_D}\}$ such that $Y_\text{train}=RF(X_\text{train})$ where $RF(X)$ is the trained random forest.
The trained random forest then serves as the ridesharing cost function such that $g(\mathbf{m}_c) = -RF(\mathbf{m}_c)$ where the minus sign is included to maximize customer rating.
It is assumed that a customer's 5-star rating is given purely to reflect their ride metrics and not factors such as driver interactions which are not necessary for autonomous MOD systems.

%% file: experiments_arxiv.tex
The predictive positioning and ridesharing methods are tested using simulation of the MIT MOD system.
Specially, there are two motivating test cases: 1) evaluating how service times for customers are affected under different fleet management strategies as customer arrival rates grow; and 2) evaluating how customer QoS is affected by various ridesharing strategies operating under a range of unknown customer preference models.
The MIT MOD system is used to provide simulation parameters that reflect a realistic operating environment for vehicles and customers.

\subsection{Simulation Setup}
\begin{figure}
	\center
	\includegraphics[width=1\columnwidth]{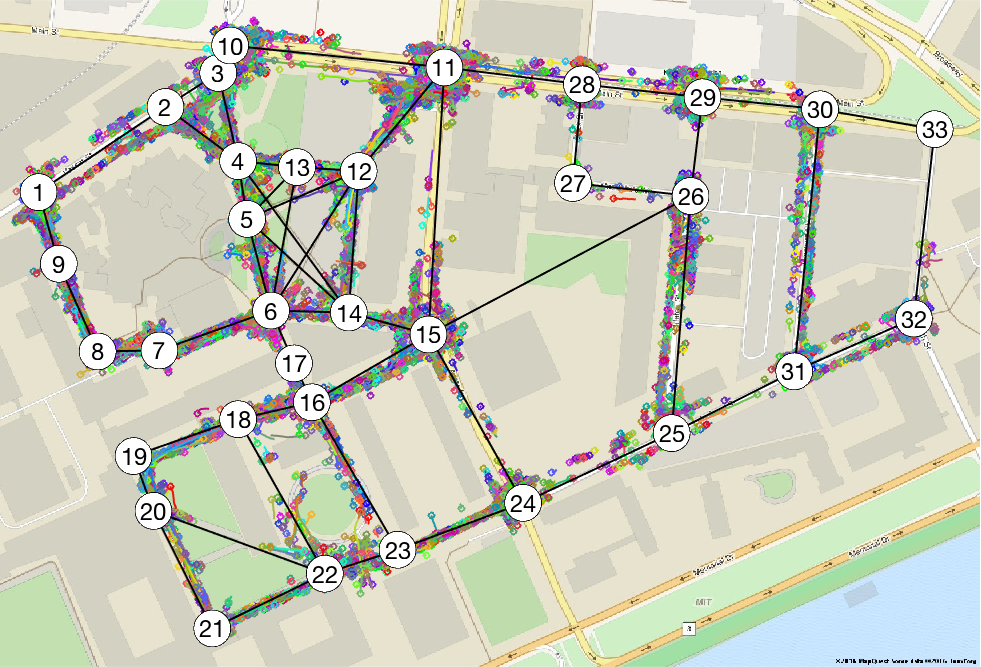}
	\caption{Pedestrian traffic network on MIT campus with overlaid pedestrian trajectories. The network graph is composed of 27 nodes, 106 directed links, and 1056 precomputed routes.}
	\label{fig::network_graph}
\end{figure}

Pedestrians and vehicles operate within a network graph for the MIT campus.
The network graph, shown in \cref{fig::network_graph}, is generated using pedestrian trajectory data collected from sensors onboard the MOD vehicles following the method presented in \cite{miller_dynamic_2016}.
A two hour time period is simulated; during which time a subset of 10 randomly chosen nodes are assigned non-zero pedestrian arrival rates in order to reflect that not every campus location experiences arrivals at all times. 
Customer arrival rates are static for the time period and take values between 0 and 0.45 ped/min at each node.
There are 3 vehicles in the MIT MOD system each with a maximum capacity of 3 passengers.
Vehicles travel between nodes according to the schedule generated by the ridesharing algorithm.
A vehicle picks up its assigned customers upon reaching a scheduled node.
If a vehicle's schedule is empty, the vehicle will travel to nodes prescribed by the predictive positioning algorithm.
Vehicle link speeds are either 11m/s for links corresponding city streets or 4m/s for campus pathways.

\subsection*{Predictive Positioning}
\begin{figure}
	\center
	\includegraphics[width=1\columnwidth]{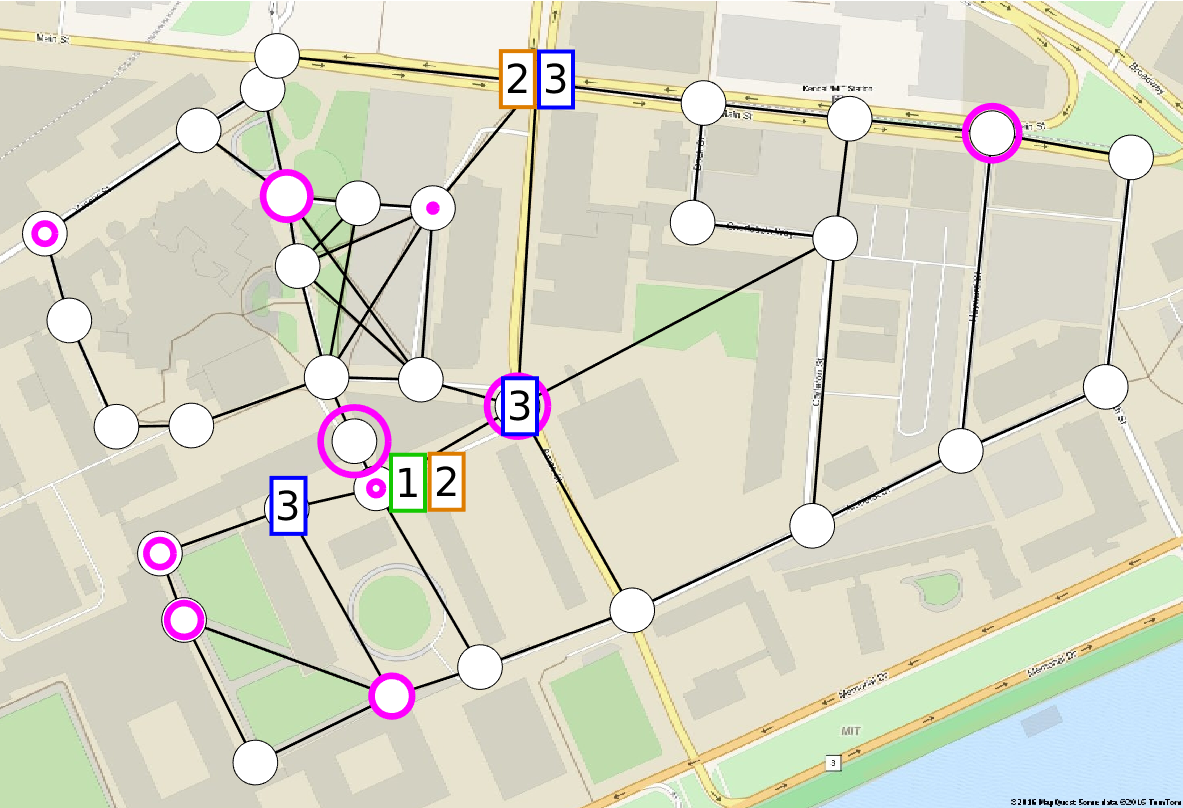}
	\caption{Relative customer arrival rates and computed predictive nodes. The numbered predictive nodes illustrate where unassigned MOD vehicles should be located to minimize the expected wait time for customers. The radius of the circles indicates the relative magnitudes of the node arrival rates. The numbers indicate the locations to place either 1, 2, or 3 vehicles depending on the current number of unallocated vehicles. }
	\label{fig::predictive_positions}
\end{figure}

The predictive positioning algorithm finds key predictive node positions to place unallocated MOD vehicles based on customer arrival rate parameters.
\Cref{fig::predictive_positions} shows the predictive nodes chosen for either 1, 2, or 3 unallocated vehicles for a particular set of arrival rates.
The predictive nodes take into account both the probability of the arrivals occurring and the vehicles' travel time to reach each node.
If only a single vehicle is unallocated, it will tend to be positioned centrally within the network, but skewed towards large arrival rates. 
When more vehicles are unallocated, the predictive nodes are further spread out for better coverage. 

The performance of the predictive positioning algorithm is evaluated through comparison against a baseline unmanaged MOD strategy, where vehicles respond to pickup requests but are not repositioned after dropping off customers.
The two methods are first compared without the use of ridesharing (vehicle capacities are 1), with assignments evaluated using a minimum service time cost function.
\Cref{fig::methods_all} shows that predictive positioning is able to reduce customer service times.
When arrival rates are lower than 0.35 ped/min per vehicle, there is often time between arrivals for vehicles to reposition to the predictive nodes and service times can be reduced by up to 20\%.
As arrival rates increase, however, the benefits of predictive positioning are reduced as vehicles are continually allocated to requests.
Under high arrival rates, it would be desirable to add more vehicles to shift to the lower portion of the curve.
However, with a fixed-sized fleet, that option is not available so ridesharing is used as an alternative.

\subsection{Ridesharing}
\begin{figure}[t]
	\center
	\includegraphics[width=1\columnwidth]{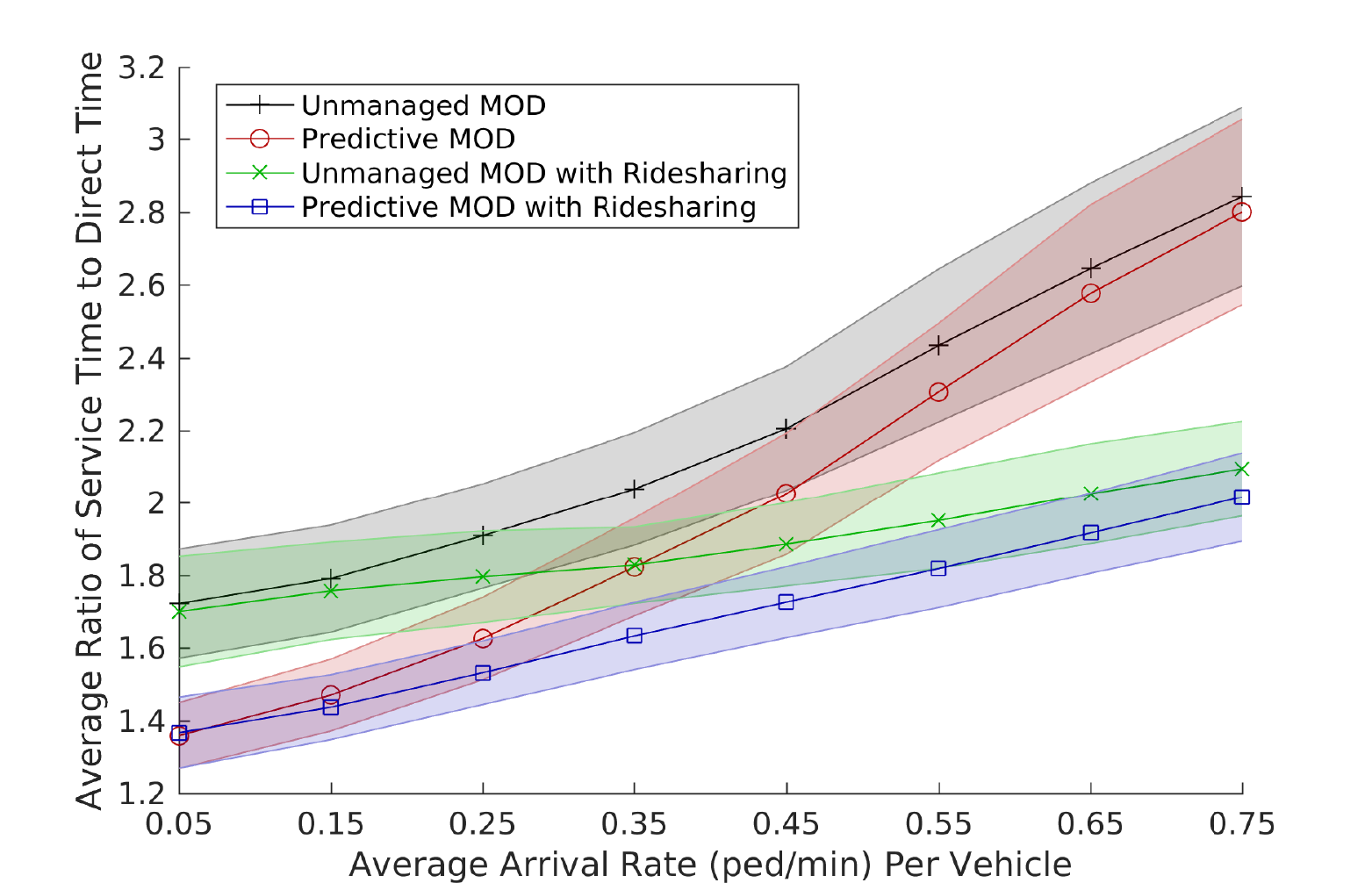}
	\caption{Service times for unmanaged and predictive positioning methods, with and without ridesharing, over a range of customer arrival rates. The figure shows that predictive positioning reduces service times when arrival rates are low and ridesharing reduces service times when arrival rates are high.  The service times are normalized by the direct time so as not to penalize service times for customers with longer routes. Lower service times are better. The arrival rates are normalized by the number of vehicles. The mean and standard deviation from 100 iterations are shown.}
	\label{fig::methods_all}
\end{figure}

The performance of the ridesharing algorithm is evaluated by comparing the single capacity predictive positioning and unmanaged MOD methods with ridesharing versions where the maximum vehicle capacity is increased to 3.
The ridesharing algorithm is applied using a minimum service time cost function.
\Cref{fig::methods_all} also shows that ridesharing reduces customer service times.
When arrival rates are lower than 0.35 ped/min per vehicle, the ridesharing methods perform similarly to their single capacity counterparts since current customers can be serviced before new customers arrive.
But at higher arrival rates, the ridesharing algorithm begins to utilize the excess vehicle capacity and new customers are inserted into vehicle schedules before previous customers have finished their ride.
Through the use of a combined predictive positioning and ridesharing approach, the MOD system is able achieve a better customer QoS across all arrival rates, resulting in as much as a 29\% reduction in service time compared to the single capacity unmanged MOD strategy.

\subsection{Customer Preference}
In the previous analysis, the customer preference was assumed to be focused on service time.
However, this assumption can be wrong and customers may give poor ratings if the the true customer preference lies elsewhere in other QoS metrics.
To evaluate the rating performance of an MOD system, a simulated rating model is used to assign a 5-star ratings to customers based on a set of customer preference weights.
The details of the simulated rating model are provided in \cref{sec::rating_model}, which also specifies how customer preference is encoded using weight concentration parameters.
Six customer preference modes are analyzed, where the weight concentration parameters are  90\% skewed towards either wait time, ride time, service time, the number of stops while onboard, ride distance, or a combined weight between service time and ride distance.
Five fleet management strategies are considered.
First, a minimum vehicle distance strategy is considered, where assignments are not made based on customer ratings but rather based on the traditional minimum vehicle travel distance metric which would minimize fuel consumption.
Next, three focused strategies based on ride time, service time, and wait time are included, where each strategy is given access to the underlying ground truth ratings function but chooses rating weights according to its focus.
Finally, the presented random forest ratings model strategy is included where the ground truth ratings function is not available but rather customer preference is learned from 5-star customer feedback ratings.
The random forest model is implemented using \cite{jaiantilal_randomforest-matlab_????}, which is trained separately under each customer preference mode for 10 runs.
To further test the ratings model, all ride distance metrics are removed from the random forest feature vector in order to see if performance can be learned using only non-corresponding, but correlated metrics.
The arrival rate is fixed at 0.35 ped/min per vehicle so that the MOD fleet is operating under the ridesharing regime.

\begin{figure}
	\centering
	\begin{subfigure}{0.24\textwidth}
		\includegraphics[width=1\columnwidth]{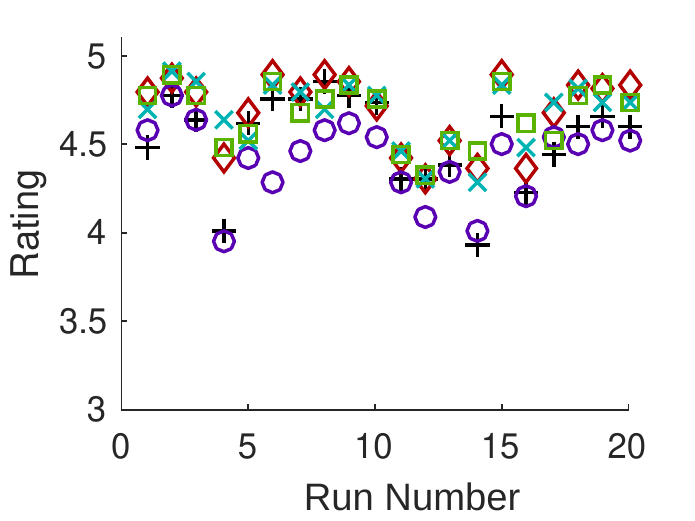} %\vspace*{-.2in}
		\caption{Wait Time}
		\label{fig::wait_time}
	\end{subfigure} \hspace*{-.1in}
	\begin{subfigure}{0.24\textwidth}
		\includegraphics[width=1\columnwidth]{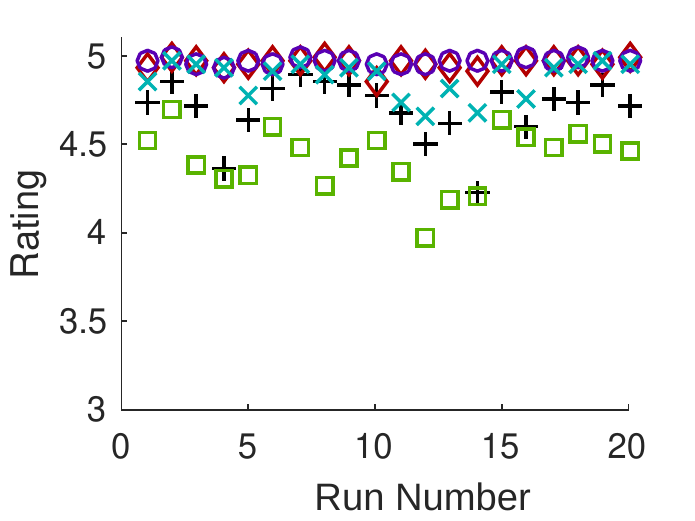}
		\caption{Ride Time}
		\label{fig::ride_time}
	\end{subfigure}
	\begin{subfigure}{0.24\textwidth}
		\includegraphics[width=1\columnwidth]{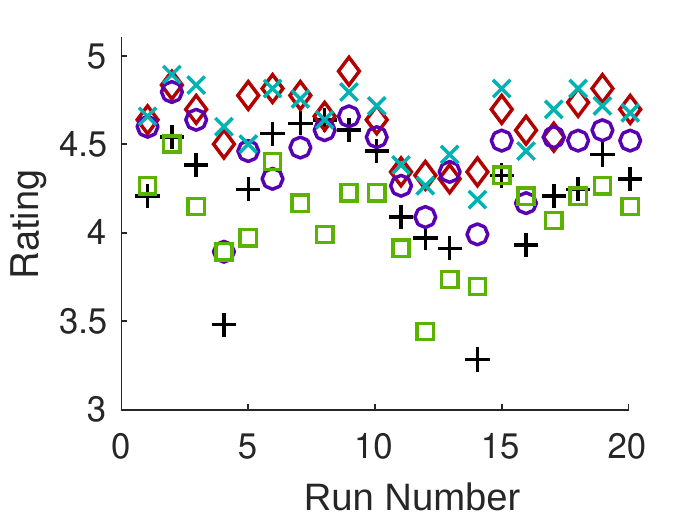}
		\caption{Service Time}
		\label{fig::service_time}
	\end{subfigure} \hspace*{-.1in}
	\begin{subfigure}{0.24\textwidth}
		\includegraphics[width=1\columnwidth]{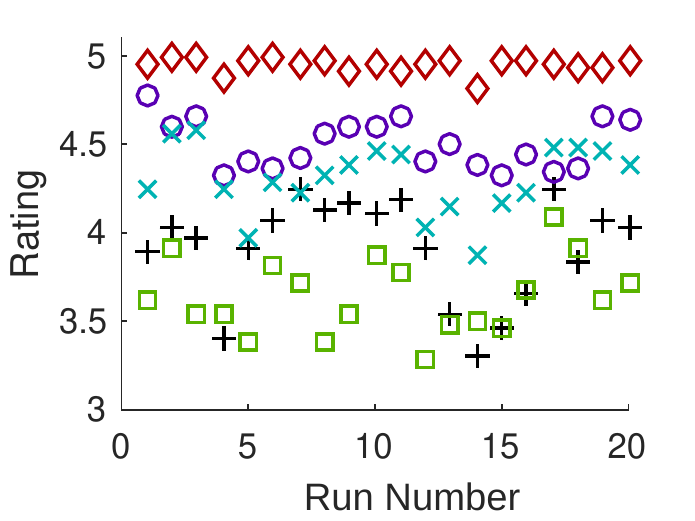}
		\caption{Number of Stops}
		\label{fig::number_of_stops}
	\end{subfigure}
	\begin{subfigure}{0.24\textwidth}
		\includegraphics[width=1\columnwidth]{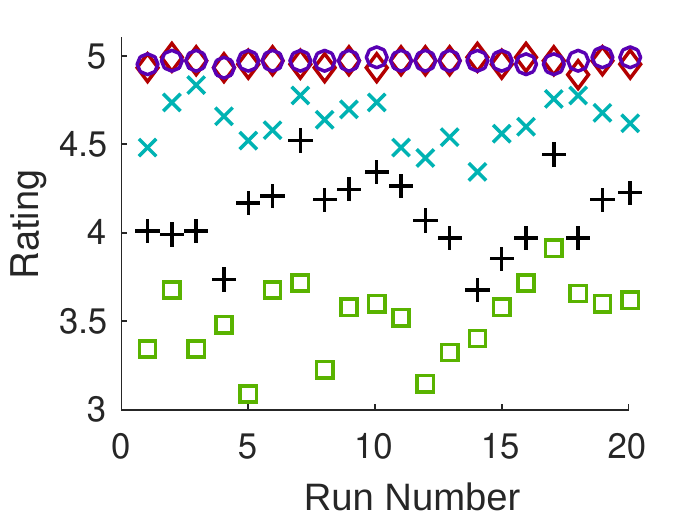}
		\caption{Excess Ride Distance}
		\label{fig::excess_ride_distance}
	\end{subfigure} \hspace*{-.1in}
	\begin{subfigure}{0.24\textwidth}
		\includegraphics[width=1\columnwidth]{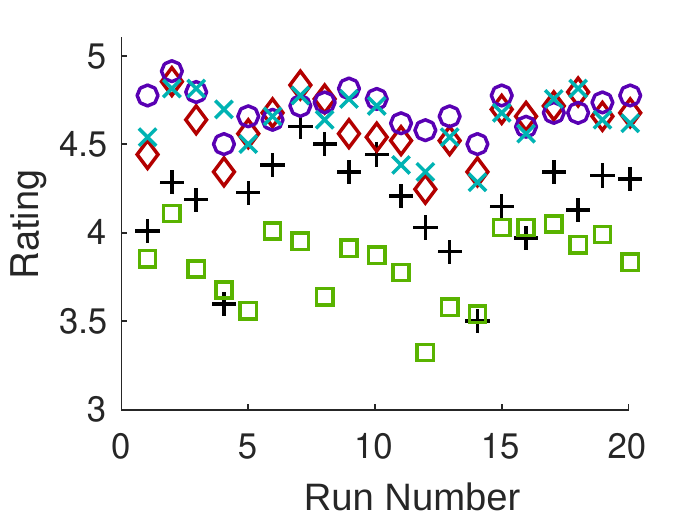}
		\caption{Combined}
		\label{fig::combined}
	\end{subfigure}
	\begin{subfigure}{0.3\textwidth}
		\vspace{5pt}
		\includegraphics[width=1\columnwidth]{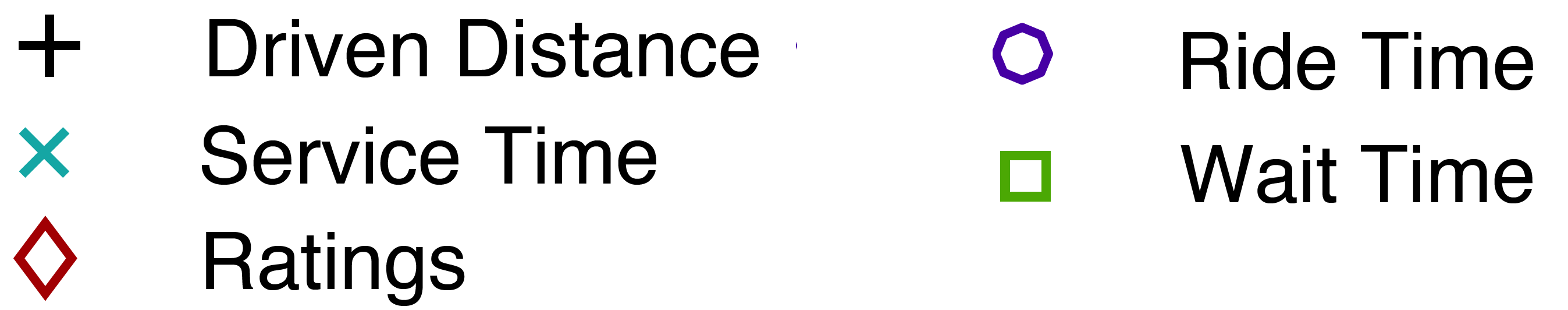}
	\end{subfigure}
	\vspace{5pt}
	\caption{Average customer rating for several fleet management strategies under customer preferences that are skewed towards either (a) wait time, (b) ride time, (c) service time, (d) the number of stops while onboard, (e) the excess ride distance between the received ride and a direct ride, or (f) a combined skew between service time and excess ride distance. The figures show that the performance of each strategy is dependent on the underlying customer preference, with the exception of the ratings strategy which performs well in all cases. Each figure shows the results of 20 runs. Performance is measured by the average customer rating where higher ratings are better. }
	\label{fig:ratings_results}
\end{figure}

\Cref{fig:ratings_results} shows how the performance, in terms of average received rating, for each strategy depends on the underlying customer preferences.
\Cref{tab::ratings} summarizes the average customer rating over all 20 iterations.
The results show that wait time, ride time, and service time each perform best when the customer preference mode matches.
Additionally, the ride time metric performs best under the excess ride distance and combined customer preference modes because ride distance and ride time are correlated.
However, each of the focused fleet management strategies performs relatively poorly under at least one customer preference mode, and the minimum distance strategy always performs poorly because customer preference is not considered.
In contrast, the ratings model demonstrates robust performance across all customer preference modes. 
The ratings model is within 0.5\% of the focused strategies under their respective modes, demonstrating that it was able to learn which customer metrics were important under each mode.
The excess ride distance mode illustrates how the ratings model is able to learn customer preference using elements in its feature vector that are only correlated with the customer preference metric and not included directly.
Finally, when considering the average performance over all customer preference modes, the ratings model performed best.

\begin{table}[t]
	\centering
	\setlength\tabcolsep{2pt}
	\caption{MOD fleet management performance as measured by average customer rating.}
	\begin{tabular}{c || c c c c c c | c}
		{} & \multicolumn{5}{c}{Customer Preference} \\
		Strategy & Wait & Ride & Service & \# Stops & Distance & Combined & Average \\
		\hline \hline 
		Distance & 4.53 & 4.70 & 4.22 & 3.91 & 4.10 & 4.17 & 4.27 \\ 
		Ride Time & 4.42 & \textbf{4.97} & 4.43 & 4.50 & \textbf{4.97} & \textbf{4.70} & 4.66  \\ 
		Service Time & 4.67 & 4.88 & \textbf{4.64} & 4.30 & 4.62 & 4.63 & 4.62 \\
		Wait Time & \textbf{4.69} & 4.42 & 4.09 & 3.64 & 3.51 & 3.82 & 4.03 \\		
		Ratings & 4.67 & 4.96 & 4.63 & \textbf{4.95} & 4.96 & 4.60 & \textbf{4.80}  \\ 
	\end{tabular} 
	\label{tab::ratings}
	\vskip -0.1in
\end{table}

%% file: conclusion.tex
This work demonstrated that predictive positioning and ridesharing can be utilized in an MOD system in order to improve customer QoS.
A predictive positioning algorithm was presented which uses customer arrival rate information to position vehicles at key nodes in the MOD network graph which minimize the expected customer wait time. 
In a simulated campus setting, the predictive positioning method was shown to reduce customer service times by as much as 20\% when customer arrival rates are low.
To improve QoS as arrival rates increase, a ridesharing approach was presented which utilizes a customer QoS based cost function. 
A combined predictive positioning and ridesharing approach was shown to reduce customer service times by as much as 29\%.
A customer ratings model was introduced as a means for learning customer preference through feedback in the form of a 5-star rating. 
The customer ratings model is shown to provide the best overall MOD fleet management performance over a range of customer preferences.

The predictive positioning and ridesharing methods which were evaluated for the campus setting could also be applied to larger networks. 
To improve the scalability of \Cref{alg::predictive_positioning}, predictive node locations could represent larger regions within an MOD service area, where customer wait times include both the travel time between regions and the expected travel time within a region.
The ridesharing method can be applied to larger fleets by parallelizing the scheduling computations for each vehicle across multiple machines which bid to a centralized machine for the customer allocations. 

Future work will apply these methods to the physical MIT MOD system. 
Methods for learning and predicting customer arrival rate trends will be studied so that vehicles can be predictively positioned to match real time demand.
Additionally, ratings and service metrics from actual customers will be used to improve and evaluate rating models based on random forest and other machine learning techniques.

%% file: appendix.tex
\subsection{Computing Customer Ride Metrics} \label{sec::ride_metrics}
This section provides details on how customer metrics are evaluated using vehicle schedule information. 
Customer $c$ makes a request at the point in time $\hat{t}^{\,\text{request}}_c$ and is assigned to $v$ currently located at node $n_v$. 
The pickup $p_c$ and drop off $e_c$ nodes for $c$ occur at respective nodes $s_i$ and $s_j$ in the vehicle schedule $\mathbf{s}_v$.
The vehicle travels between any adjacent nodes $s_k$ and $s_{k+1}$ in its schedule using route $r(s_{k},s_{k+1})$. 
The travel distance and travel time between the nodes are
\begin{align}
d(s_{k},s_{k+1}) &= \sum_{l \in \mathcal{L}_{r(s_{k},s_{k+1})}} d_l , \\
t(s_{k},s_{k+1}) &= \sum_{l \in \mathcal{L}_{r(s_{k},s_{k+1})}} \frac{d_l}{u_l} \label{eqn::travel_time} ,
\end{align}
where $d_l$ and $u_l$ are the length and average travel speed of link $l$, respectively.

The metrics $\mathbf{m}_c^{vij}$ are computed as follows:
\begin{align}
\hat{t}^{\,\text{pickup}}_c &= t(n_v,s_{1}) + \sum_{k=1}^{i-1} t(s_{k}, s_{k+1}), \\
\hat{t}^{\,\text{dropoff}}_c &= t(n_v,s_1) + \sum_{k=1}^{j-1} t(k_{k}, s_{k+1}),\\
t^{\text{direct}}_c &= t(p_c, e_c), \\
d^{\text{direct}}_c &= d(p_c, e_c), \\
t^{\text{walk}}_c &= \bar{t}(p_c, e_c), \\
t^{\text{ride}}_c &= \hat{t}^{\,\text{dropoff}}_c - \hat{t}^{\,\text{pickup}}_c, \\
t^{\text{wait}}_c &= \hat{t}^{\,\text{pickup}}_c - \hat{t}^{\,\text{request}}_c,\\
t^{\text{service}}_c &= t^{\text{wait}}_c + t^\text{ride}_c, \\
t^{\text{ratio}}_c &= t^\text{ride}_c/t^\text{direct}_c, \nonumber \\
t^{\text{excess\_ride}}_c &= t^\text{ride}_c - t^\text{direct}_c, \\
N^{\text{stops}}_c &= k-j, \\	
t^\text{notify}_c &= \hat{t}^{\,\text{assigned}}_c - \hat{t}^{\,\text{request}}_c, \\
d^{\text{traveled}}_c &= \sum_{k=i}^{j-1} d(s_{k}, s_{k+1}), \\
t^\text{excess\_walk}_c &= t^\text{service}_c - t^\text{walk}_c, 
\end{align}
where $\hat{t}^{\,\text{pickup}}_c$ is the point in time $c$ is picked up, $\hat{t}^{\,\text{dropoff}}_c$ is the point in time $c$ is dropped off, $t^{\text{direct}}_c$ is the time it would take to drive directly from $p_c$ to $e_c$, $d^{\text{direct}}_c$ is the direct route distance between $p_c$ and $e_c$, $t^{\text{walk}}_c$ is the time it would take to walk from $p_c$ to $e_c$, $\bar{t}(n_i,n_j)$ is \Cref{eqn::travel_time} evaluated with $r(n_i,n_j)$ and $v_l$, as the respective route and speed of the pedestrian instead of a vehicle, and $\hat{t}^{\,\text{assigned}}_c$ is the point in time when the ridesharing algorithm assigns the customer to the vehicle.
Note, if a customer is rejected ($1^\text{rej}_c = 1$), then many of the metrics do not apply and the customer metrics are set to be $\mathbf{m}_c^\text{rejected} = \{1^\text{rej}_c, t^{\text{notify}}_c, t^{\text{walk}}_c \}$.
If the customer is given a ride, then the metrics are $\mathbf{m}_c = \{1^\text{rej}_c, t^{\text{ride}}_c, t^{\text{wait}}_c, t^{\text{service}}_c, t^{\text{ratio}}_c, t^{\text{excess\_ride}}_c, N^{\text{stops}}_c, \allowbreak t^{\text{notify}}_c, d^{\text{traveled}}_c, t^{\text{walk}}, t^{\text{excess\_walk}}_c \}$.

\subsection{Simulated Rating Model} \label{sec::rating_model}
This section presents a simulated rating model, which is used as ground truth in simulation to assign 5-star ratings to MOD customers.
The values and functional forms are chosen based on an assumed customer preference and are kept hidden from the 5-star rating model.

If a customer is rejected, then they give one of the two lowest ratings based on how long they waited to be notified of their rejection. 
A rejected customer's rating is
\begin{equation}
r^\text{rejected} = 
\begin{cases}
2, & \text{if} \quad \frac{t^{\text{notify}}_c}{t^{\text{walk}}_c} \le 0.1 \\
1, & \text{otherwise.}
\end{cases}
\label{eqn::rejected_rating}
\end{equation}

If a customer is given a ride, then the rating will be a weighted sum of 5 aggregate ratings based on wait time, ride time, service time, number of stops, and ride distance computed as 
\begin{equation}
r^\text{accepted}_c = w_1 r^\text{wait}_c + w_2 r^\text{ride}_c + w_3 r^\text{service}_c + w_4 r^\text{stops}_c + w_5 r^\text{distance}_c ,
\label{eqn::accepted_rating}
\end{equation}
with
\begin{align*}
r^\text{wait}_c &= \text{Range}\left(\frac{t^{\text{wait}}_c}{t^{\text{walk}}_c - t^{\text{direct}}_c},0,1\right)  \\
r^\text{ride}_c &= \text{Range}\left(\frac{t^{\text{ride}}_c - t^{\text{direct}}_c}{t^{\text{walk}}_c - t^{\text{direct}}_c},0,1\right)  \\
r^\text{service}_c &= \text{Range}\left(\frac{t^{\text{service}}_c - t^{\text{direct}}_c}{t^{\text{walk}}_c - t^{\text{direct}}_c},0,1\right)  \\
r^\text{stops}_c &=  \text{max}(1, 6 - N^{\text{stops}}_c) \\
r^\text{distance}_c &= \text{Range}\left(\frac{d^{\text{traveled}}_c - d^{\text{direct}}_c}{d^{\text{direct}}_c},0,0.5\right) ,\\
\end{align*}
where $\text{Range}(\alpha,\beta,\gamma)$ maps $\alpha$ to the $i$-th interval of 5 exponentially spaced values between $\beta$ and $\gamma$ and assigns the value 6-$i$ as the rating. Range values were chosen to reflect a set of possible expected customer satisfaction levels for each metric. For example, setting the $\gamma$ value for $r^\text{distance}_c$ to 0.5 reflects that customers would give the lowest rating once their journey distance exceeded the nominal distance by a factor of 0.5. Exponential spacing is used to cause ratings to drop off more quickly as metrics worsen for customers. 

The majority of customers follow the same set of weights $\mathbf{w}$, but some customers do not.
To accommodate this, the weights are drawn from a Dirichlet distribution such that $w \sim \text{Dir}(\hat{w})$, where the concentration parameters $\hat{w}$ represent the nominal weights for the population.